\documentclass[%
 reprint,,bibtex,
superscriptaddress,
nofootinbib,
 amsmath,amssymb,
 aps,
prb
]{revtex4-1}
\usepackage{pdfpages}
\usepackage{units}
\usepackage{graphicx}% Include figure files
\usepackage{dcolumn}% Align table columns on decimal point
\usepackage{bm}% bold math
\begin{document}

\preprint{APS/123-QED}

\title{ Magnetic Proximity Effect and Interlayer Exchange Coupling of
Ferromagnetic/Topological Insulator/Ferromagnetic Trilayer
}

\author{Mingda Li}
\email{mingda@mit.edu}
\affiliation{Department of Nuclear Science and Engineering, Massachusetts Institute of Technology, 77 Massachusetts Avenue, Cambridge, MA 02139, USA}%Lines break automatically or can be forced with
\affiliation{Fracsis Bitter Magnetic Lab, Massachusetts Institute of Technology, 77 Massachusetts Avenue, Cambridge, MA 02139, USA}

\author{Wenping Cui}
\affiliation{Institut f\"ur Theoretische Physik, Universit\"at zu K\"oln, Z\"ulpicher Str. 77, D-50937 K\"oln, Germany}

\author{Jin Yu}
\affiliation{Department of Nuclear Science and Engineering, Massachusetts Institute of Technology, 77 Massachusetts Avenue, Cambridge, MA 02139, USA}
\affiliation{State Key Laboratory of Mechanics and Control of Mechanical Structures, Nanjing University of Aeronautics \& Astronautics, Nanjing 210016, China}

\author{Zuyang Dai}%
\affiliation{%
Department of Physics, Tsinghua University, Beijing 100084, China
}%

\author{Zhe Wang}
\affiliation{Department of Nuclear Science and Engineering, Massachusetts Institute of Technology, 77 Massachusetts Avenue,
Cambridge, MA 02139, USA}%Lines break automatically or can be forced with \\

\author{\\Ferhat Katmis}
\affiliation{Fracsis Bitter Magnetic Lab, Massachusetts Institute of Technology, 77 Massachusetts Avenue, Cambridge, MA 02139, USA}
\affiliation{Department of Physics, Massachusetts Institute of Technology, 77 Massachusetts Avenue, Cambridge, MA 02139, USA}

\author{Wanlin Guo}
\affiliation{State Key Laboratory of Mechanics and Control of Mechanical Structures, Nanjing University of Aeronautics \& Astronautics, Nanjing 210016, China}

\author{Jagadeesh Moodera}
\email{moodera@mit.edu}
\affiliation{Fracsis Bitter Magnetic Lab, Massachusetts Institute of Technology, 77 Massachusetts Avenue, Cambridge, MA 02139, USA}
\affiliation{Department of Physics, Massachusetts Institute of Technology, 77 Massachusetts Avenue, Cambridge, MA 02139, USA}

\date{\today}% It is always \today, today,

\begin{abstract}
Magnetic proximity effect between topological insulator (TI) and ferromagnetic insulator (FMI) is
considered to have great potential in spintronics. However, a complete determination of interfacial
magnetic structure has been highly challenging. We theoretically investigate the interlayer exchange
coupling of two FMIs separated by a TI thin film, and show that the particular electronic states of the TI
contributing to the proximity effect can be directly identified through the coupling behavior between two
FMIs, together with a tunability of coupling constant. Such FMI/TI/FMI structure not only serves as a platform to clarify the magnetic structure of FMI/TI interface, but also provides insights into designing the magnetic storage devices with ultrafast response.
%\begin{description}
%\end{description}
\end{abstract}
\maketitle

%\tableofcontents

\section{Introduction}
The breaking of time-reversal symmetry (TRS), which opens up a gap to the helical Dirac surface states in a three-dimensional strong topological insulator (TI), has been shown to be of central importance for both fundamental aspects \cite{1luo2013massive, 2li2010dynamical, 3qi2009inducing,
 4liu2009magnetic, 5qi2008topological, 6qi2011topological, 7hasan2010colloquium} and device applications \cite{2li2010dynamical, 6qi2011topological, 7hasan2010colloquium,8Mlang,9li2012thz,10wray2012device,11vobornik2011magnetic} in TI studies.  For instance the topological magnetoelectric effect \cite{7hasan2010colloquium}, which enables the possibility of electric-field controlled spin transistor \cite{10wray2012device,12xue2011nanoelectronics}, requires an opening of surface band gap as a prerequisite to reach ``off'' state, otherwise the gapless surface state  would lead to a leakage current and very low on/off ratio. Another promising example is the realization of quantum anomalous Hall effect \cite{13chang2013experimental,14xue2014preface,15wang2013quantum,16yu2010quantized}, where gapped surface states are accompanied with backscattering-protected edge transport channels without external magnetic field. This opens up the possibility for developing next generation low-dissipation spintronic devices. Moreover, domain-wall Majorana bound states are predicted at the TI/FMI interface where magnetization switches sign, which could be applied in error-tolerant topological quantum computation \cite{6qi2011topological, 7hasan2010colloquium, 17bernevig2013topological}.  All the examples above require a gap-opening of the surface states of TI.

In general, there are two approaches to break the TRS and open up the gap: magnetic doping \cite{13chang2013experimental,14xue2014preface,15wang2013quantum,18men2011carrier} and magnetic proximity effect \cite{8Mlang,14xue2014preface,19wei2013exchange,20men2014engineering,21eremeev2013magnetic,22eremeev2013magnetic,23men2013interface,24men2012bound}. Compared with the doping method, the advantages of the latter include better controllability of the electronic states, uniformly distributed band gap in space, and preservation of the TI's original crystalline structure, etc. In this regard, a comprehensive understanding of the interfacial magnetic structure between TI and FMI becomes essential for observing above predicted phenomena.

However, a complete determination of the magnetic structure between TI and FMI turns out to be nontrivial. On the one hand, the interaction between TI  and the FMI states is self-consistent in nature, where TI states can lead to complex spin structure, such as magnetic precession in FMI \cite{25li2014controlling}. On the other hand, it is also hindered by the insufficient information experiments can extract for a comprehensive understanding. For instance, despite the powerful technique of spin-resolved ARPES to study surface electronic and magnetic structure of doped TI \cite{26hsieh2009observation}, due to the small ($\sim\unit[1]{nm}$) escape depth of the photoelectrons \cite{27wray2013chemically}, ARPES renders to be inapplicable to study interfacial magnetic structure where FMI layer is epitaxially grown on the top of TI. Magneto-optical Kerr effect (MOKE) is another promising method, which could be used to determine both in-plane and out-of-plane magnetization, and has been successfully applied in TI studies \cite{8Mlang}. However, the resulting signal of rotated polarization is indeed an overall effect of total magnetization projection, without sensitivity to individual layer. Due to the short-range nature of exchange coupling \cite{28bruno1999theory,29levy1998range,30bruno1996interlayer,31bruno1995theory,32bruno1993geometrical,33bruno1993interlayer,34bruno1992ruderman}, only the thin layer of magnetic moments very close to the interface contributes to the proximity effect, instead of the total measured magnetization as in MOKE. Actually, at the interface of TI/FMI structure, strong spin-orbit coupling may tilt the interfacial magnetic moment and result in a different magnetic structure near the interface \cite{19wei2013exchange,35bruno1989tight}. Another powerful characterization tool is polarized neutron reflectometry (PNR), which has shown great advantages \cite{36spurgeon2013thickness} thanks to both compositional and depth sensitivity, but PNR only measures the in-plane magnetization component, without resolving the electronic states of TI which participate in the proximity exchange coupling. Therefore, a deeper understanding of TI/FMI proximity, which considers only the near-interface FMI states, with distinguishability of particular TI states involved in the exchange-coupling process, is clearly needed.

\begin{figure}
\includegraphics[width=0.48\textwidth]{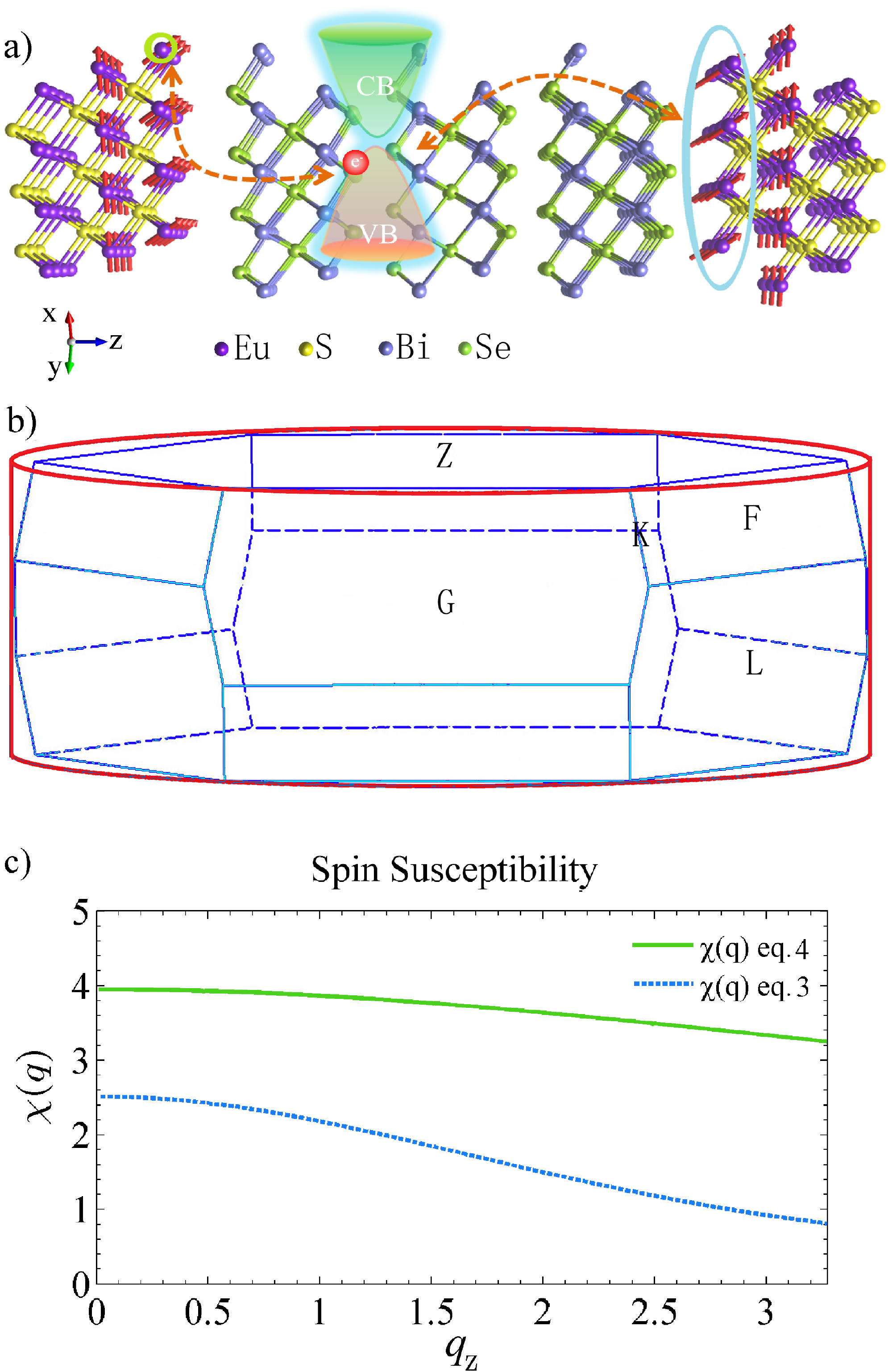}
\caption{a) The atomic configuration of EuS/Bi$_2$Se$_3$/EuS trilayer, which is a viable example for FMI/TI/FMI structure. The magnetic moment of Eu atoms are shown as red arrows. The spin structure near the interface may be canted near the interface. The interlayer coupling (orange dashed line) is achieved through the electronic states of the TI spacer. b) the original 1st Brillouin zone and the simplified cylindrical integration volume. c)The comparison betweens spin susceptibilities  using eq.(\ref{eq3}) and eq.(\ref{eq4}), with 4-band Hamiltonian.}
\label{fig1}
\end{figure}

In this study, we calculate the magnetic interlayer coupling constant between two thin layers of FMI, separated by a thin spacer layer of three-dimensional TI, within linear response theory \cite{37white2007quantum} and RKKY interactions \cite{31bruno1995theory, 34bruno1992ruderman}. We take the EuS/Bi$_2$Se$_3$/EuS as an example of FMI/TI/FMI trilayer. Since the interlayer coupling and magnetic proximity has the same origin of exchange coupling between FMI and TI, the information about the electronic states of TI which participate in the interlayer coupling process should also be responsible for the magnetic proximity effect.

We use one atomic layer thickness of magnetic moment of FMI to describe the short-range exchange coupling interaction, and apply model Hamiltonian of Bi$_2$Se$_3$ as \cite{1luo2013massive,38liu2010model} as the prototype of TI. We show that a ferromagnetic-antiferromagnetic oscillatory coupling also exists when varying the number of quintuple layer (QL) of TI, similar to the interlayer exchange coupling results in Fe/Cr/Fe \cite{31bruno1995theory, 39BentPhd05}. Most importantly, we show distinct behaviors of coupling between massive Dirac TI state and the $p_z$ bands of Bi and Se, due to the paramagnetic nature of Dirac surface state and large diamagnetism of Bi \cite{40buot1972theory,41hebborn1964conduction} orbitals. The sign difference and the tunability of coupling constant vs Fermi level can be applied independently to identify the TI states contributing to the proximity effect, due to the same origin of short-range magnetic exchange coupling process. Our approach, when applied to various FMI/TI/FMI systems, can be used to better understand the TI/FMI proximity effects, and thus for optimized designing of TI-based spintronic devices.

%%%%%%%%%%%%%%%%%%%%%%%%%%%%%%%%%%%%%%%%%%%%%%%%%%%%%%%%%%%%%%%%%%%%%%%%%%%%%%%%%%%%%%%%%%%%%%%%%%%%%%%%%%%%%
\section{Theory}
%%%%%%%%%%%%%%%%%%%%%%%%%%%%%%%%%%%%%%%%%%%%%%%%%%%%%%%%%%%%%%%%%%%%%%%%%%%%%%%%%%%%%%%%%%%%%%%%%%%%%%%%%%%%%
\subsection{Interlayer Magnetic Coupling Constant}
The FMI/TI/FMI trilayer EuS/Bi$_2$Se$_3$/EuS is schematically represented in Fig. \ref{fig1}a. For a given localized magnetic ion (Eu ion in green circle) of FMI close to the interface, the interlayer magnetic coupling constant $I_{12}$ is an overall effect of the indirect exchange coupling of all the Eu ions (blue ellipse) on the other side of TI/FMI interface, through the coupling of electronic states in TI (orange dashed lines). Due to the localized nature of Eu moments, we could apply the RKKY type of interlayer coupling strength\cite{31bruno1995theory,33bruno1993interlayer,34bruno1992ruderman},

\begin{equation}\label{eq1}
{{I}_{12}}\!=\!\!-\tfrac{A^2S^2d}{2V_0^2(2\pi)^3}\!\!\int{d{{q}_{z}}{{{d}^{2}}q_{\parallel }}{{e}^{i{{q}_{z}}z}}\!\chi ({{q}_{\parallel }},{{q}_{z}})\!\!\sum\limits_{{{R}_{\parallel }}\in {{F}_{2}}}{{{e}^{i{{q}_{\parallel }}\cdot {{R}_{\parallel }}}}}}
\end{equation}

where $A$ is the amplitude of the contact potential $A{{\overset{\scriptscriptstyle\rightharpoonup}{S}}_{i}}\cdot \overset{\scriptscriptstyle\rightharpoonup}{s}$, with ${{\overset{\scriptscriptstyle\rightharpoonup}{S}}_{i}}$ and $\overset{\scriptscriptstyle\rightharpoonup}{s}$ are the spins of FMI and TI, respectively, $V_0$ is the atomic volume, $S$  is the spin of the FMI, for Eu$^{2+}$, $S=7/2$ at \unit[0]{K}. For finite temperature $T$, in a mean field framework we can estimate $S$ as $$S(T)=S(0)\left( 1-{{\left( {T}/{{{T}_{c}}}\; \right)}^{2}} \right)$$ for EuS $T_c=\unit[16.6]{K}$. $d$ is the distance between adjacent atomic planes in its original expression, in our present situation, due to layered structure of Bi$_2$Se$_3$, it is appropriate to take $d\sim\unit[0.96]{nm}$ which is the thickness of 1 QL, since 1 QL is the smallest coarse-grained unit for electronic properties, even though 3 QL is the unit for periodic crystalline structure; $z$ is the distance between two FMI layers, $z=(N+1)d$, where $N$ is the number of QL; $R$ is the in-plane components of the coordinates of the Eu ions to be summed up, and $\chi ({{q}_{\parallel }},{{q}_{z}})$ is the $q$-dependent magnetic susceptibility of the TI spacer. The TI states participating in the exchange coupling enter into the $\chi ({{q}_{\parallel }},{{q}_{z}})$ term, and are finally reflected in $I_{12}$. This is the theoretical basis why we could study TI/FMI proximity effect by studying interlayer coupling of FMI/TI/FMI.

In eq. (\ref{eq1}), the integration of $q$ should be performed within the 1st Brillouin zone of Bi$_2$Se$_3$ (Fig. \ref{fig1}b, blue polyhedron). However, if we define $q$ and $k$ periodically in reciprocal lattice by using periodic zone scheme instead of folded zone scheme, we could define a prismatic auxiliary zone and use the reciprocal unit cell with prismatic shape. Since the in-plane area is hexagonal and close to a circle, we further define a cylindrical integration zone which shares the same volume with the original 1st Brillouin zone (Fig. \ref{fig1}b, red cylinder), which effectively reduces the integration dimension. Finally, the interlayer coupling constant can be simplified as
\begin{equation}\label{eq2}
{{I}_{1,2}}\!=\!-\frac{1}{2}{{\left( \frac{A}{{{V}_{0}}} \right)}^{2}}\frac{{S^2d^2}}{2\pi {{V}_{0}}}\!\int\limits_{-{\pi }/{d}\;}^{+{\pi }/{d}\;}{d{{q}_{z}}\chi ({{q}_{\parallel }}=0,{{q}_{z}}){{e}^{i{{q}_{z}}z}}}
\end{equation}
Here we have used the fact that in the period zone scheme, the in-plane and out-of-plane component are decoupled; for ${{q}_{\parallel }}\ne 0$, we have $$\sum\limits_{{{R}_{\parallel }}\in {{F}_{2}}}{{{e}^{i{{q}_{\parallel }}\cdot {{R}_{\parallel }}}}}=0$$.

%%%%%%%%%%%%%%%%%%%%%%%%%%%%%%%%%%%%%%%%%%%%%%%%%%%%%%%%%%%%%%%%%%%%%%%%%%%%%%%%%%%%%%%%%%%%%%%%%%%%%%%%%%%%%
\subsection{$q$-dependent Spin Susceptibility}
In order to calculate the interlayer coupling constant $I_{12}$ in eq.(\ref{eq2}), we need the magnetic susceptibility. The spin magnetic susceptibility along direction $\mu$ ($\mu=x,y,z$) $\chi_{\mu\mu,\rm spin}$ for a generic spinor state can be written using Kubo formula as \cite{52Levy}
\begin{eqnarray}\label{eq3}
&&\quad\quad\quad\quad\!{{\chi }_{\mu\mu, \rm spin}}(\mathbf{q})\!=\!\frac{\mu _{\rm B}^{2}}{4{{\pi }^{3}}}\!\!\sum\limits_{\begin{smallmatrix}
 m,occ\\
 n,empty
\end{smallmatrix}}\int{{d}^{3}}\mathbf{k}\nonumber\\
&&{{\frac{f_0({E}_{n,\mathbf{k}})-f_0({E}_{m,\mathbf{k}+\mathbf{q}})}
{{{E}_{m,\mathbf{k}+\mathbf{q}}}-{{E}_{n,\mathbf{k}}}+i\delta }}}\left| \left \langle m,\mathbf{k}+\mathbf{q} \left|S_{\mu}\right|n,\mathbf{k}
\right \rangle \right|^2
\end{eqnarray}\\
where ${{E}_{n,\mathbf{k}}}$ denotes the eigenvalue at band number $n$ and wavevector $\mathbf{k}$, with corresponding eigenstate $|n,\mathbf{k}\rangle$, $S_{\mu}$ ($S_z=I\otimes\sigma_z$,$S_x=\tau_z\otimes\sigma_x$ and $S_y=\tau_z\otimes\sigma_y$ ) is the spin operator along direction $\mu$ and ${{\mu }_{\rm B}}$ is the Bohr magneton. The integration over $k$ is over the cylindrical integration zone in Fig. \ref{fig1}b.

When the spinor structure is absent, and within plane-wave approximation, eq. (\ref{eq3}) can be greatly simplified as \cite{37white2007quantum,42hebborn1970orbital},

\begin{equation}\label{eq4}
\!{{\chi }_{\rm spin}}(\mathbf{q})\!=\!\frac{\mu _{\rm B}^{2}}{4{{\pi }^{3}}}\!\!\sum\limits_{\begin{smallmatrix}
 m,occ\\
 n,empty
\end{smallmatrix}}{\int{\frac{f_0({E}_{n,\mathbf{k}}) - f_0({E}_{m,\mathbf{k}-\mathbf{q}})}
{{{E}_{m,\mathbf{k}-\mathbf{q}}}-{{E}_{n,\mathbf{k}}}+i\delta }}}{{d}^{3}}\mathbf{k}
\end{equation}\\

where $f_0$ is the Fermi-Dirac distribution function. \\
The comparison between eq. (\ref{eq3}) and eq. (\ref{eq4}) is shown in Fig. \ref{fig1}c. The resulting spin susceptibility (calculated using eq. (\ref{eq3}) and overlap of eigenstates of Hamiltonian eq. (\ref{eq7})), is $\sim1/2$ compared with the result using the simplified version eq. (\ref{eq4}). This could be understood as a consequence of spin texture of bands, where electronic transition amplitude for minority spin components is suppressed due to the lack of population in $\sim1/2$ of the k space\cite{54Louie}. Actually, in order to calculate the exact magnitude of susceptibility, density-functional perturbation theory method which requires the input of realistic states and summation over all bands, is needed\cite{53Louie}. However, since we are more interested in the role that TI state plays in proximity effect, in addition the effective Hamiltonian approach we adopt involves only few bands, in the following we use eq. (\ref{eq4}) instead of eq. (\ref{eq3}) to calculate the interlayer coupling constant in eq. (\ref{eq2}), for computational simplicity but without loss of qualitative illustration.

%%%%%%%%%%%%%%%%%%%%%%%%%%%%%%%%%%%%%%%%%%%%%%%%%%%%%%%%%%%%%%%%%%%%%%%%%%%%%%%%%%%%%%%%%%%%%%%%%%%%%%%%%%%%%
\subsection{Estimation of Orbital Magnetic Susceptibility }
Besides the spin susceptibility which contributes to paramagnetism, due to diamagnetic nature of bulk Bi$_2$Se$_3$, we include the diamagnetic orbital term as well. The $q$-dependent orbital susceptibility can be regarded as an overlap between eigenstates and their curvatures  \cite{41hebborn1964conduction, 42hebborn1970orbital, 53Louie}. In the $q\rightarrow0$ limit, the susceptibility from spin paramagnetism and orbital diamagnetism can be simplified as\cite{37white2007quantum}\\
\begin{equation}\label{eq5}
\chi_{\rm orb}(q\rightarrow0)=-\frac{4}{3}\left(\frac{m_e}{m^*g^*}\right)^2\chi_{\rm spin}(q\rightarrow0)\\
\end{equation}
where $m^*$ is the effective mass of electron, $D(E)$ is the density of state near energy $E$. In the case of Dirac surface state, $g^*\simeq \frac{2m_e}{m^*}$ is valid\cite{50DiracX}, hence we expect the spin paramagnetism is dominant for surface states. This is consistent with the recent experimental report about paramagnetic Dirac susceptibility in TI\cite{45zhao2014singular}. On the contrary, for bulk parabolic-like bands, $g^*\simeq2$. \cite{37white2007quantum} Due to the small effective mass of Bi$_2$Se$_3$, we expect the orbital diamagnetism dominates the spin paramagnetism in bulk Bi$_2$Se$_3$, which is also true based on experimental value \cite{43kumar2007diamagnetic}. Neglecting the Van Vleck paramagnetism which is only significant at high temperature\cite{37white2007quantum}, the total magnetic susceptibility at low temperature can be written as
\begin{equation}\label{eq6}
\chi (\mathbf{q})\simeq {{\chi }_{\rm orb}}(\mathbf{q})+{{\chi }_{\rm spin}}(\mathbf{q})
\end{equation}

%%%%%%%%%%%%%%%%%%%%%%%%%%%%%%%%%%%%%%%%%%%%%%%%%%%%%%%%%%%%%%%%%%%%%%%%%%%%%%%%%%%%%%%%%%%%%%%%%%%%%%%%%%%%%
\subsection{4-band Model Hamiltonian of TI Bi$_2$Se$_3$ }
In order to calculate the magnetic susceptibility in eq. (\ref{eq6}), eigenvalues from a model Hamiltonian are needed. Using a 4-band $k\cdot p$ theory, and a basis$ \left| p_{1z}^{+},\uparrow  \right\rangle $, $\left| p_{2z}^{-},\uparrow  \right\rangle $, $\left| p_{1z}^{+},\downarrow  \right\rangle$,
$\left| p_{2z}^{-},\downarrow  \right\rangle$,
the model Hamiltonian of a TI in Bi$_2$Se$_3$ family can be written as \cite{22eremeev2013magnetic,38liu2010model}

\begin{eqnarray}\label{eq7}
H(k)&=&{{\varepsilon }_{0}}(k){{I}_{4\times 4}}+M(k)I\otimes {{\sigma }_{z}}+{{A}_{1}}{{k}_{z}}{{\sigma }_{z}}\otimes {{\tau }_{x}}\nonumber\\
&&+{{A}_{2}}{{k}_{x}}{{\sigma }_{x}}\otimes {{\tau }_{x}}-{{A}_{2}}{{k}_{y}}{{\sigma }_{y}}\otimes {{\tau }_{x}}
% &=&{{\varepsilon }_{0}}(k){{I}_{4\times 4}}+
% \left( \begin{matrix}
%   M(k){{\sigma }_{z}}+{{A}_{1}}{{k}_{z}}{{\sigma }_{x}} & {{A}_{2}}{{k}_{-}}{{\sigma }_{x}}  \\
%  {{A}_{2}}{{k}_{+}}{{\sigma }_{x}} & M(k){{\sigma }_{z}}-{{A}_{1}}{{k}_{z}}{{\sigma }_{x}}  \\
%\end{matrix} \right)\nonumber
\end{eqnarray}

where $\varepsilon_0(k)=C+D_{1}k_z^2+D_2(k_x^2+k_y^2)$, $k_{\pm}=k_x\pm ik_y$, $M(k)=M_0-B_1k_z^2-B_2(k_x^2+k_y^2)$. For Bi$_2$Se$_3$ we have $C=\unit[-0.0068]{eV}$, $D_{1}=\unit[0.013]{eV\cdot nm^{2}}$, ${{D}_{2}}=\unit[0.196]{eV\cdot nm^{2}}$, ${{M}_{0}}=\unit[0.28]{eV}$, $B_{1}=\unit[0.10]{eV\cdot nm^{2}}$, $B_2=\unit[0.566]{eV\cdot nm^{2}}$, $A_1=\unit[0.22]{eV\cdot nm}$ and $A_2=\unit[0.41]{eV\cdot nm}$.
The doubly degenerate eigenvalues can be written as
\begin{eqnarray}\label{eq8}
\begin{cases}
\!\!E(2_{z,\uparrow /\downarrow }^{-},\!k)\!=\!{{\varepsilon }_{0}}(k)\!+\!\!\sqrt{{{M}^{2}}(k)\!+\!A_{1}^{2}k_{z}^{2}\!+\!A_{2}^{2}(k_{x}^{2}+k_{y}^{2})}\\
\!\!E(1_{z,\uparrow /\downarrow }^{+},\!k)\!=\!{{\varepsilon }_{0}}(k)\!-\!\!\sqrt{{{M}^{2}}(k)\!+\!A_{1}^{2}k_{z}^{2}\!+\!A_{2}^{2}(k_{x}^{2}+k_{y}^{2})}
\end{cases}
\end{eqnarray}
where 1=Bi and 2=Se in this notation.
\begin{figure}
\centering
\includegraphics[width=0.5\textwidth]{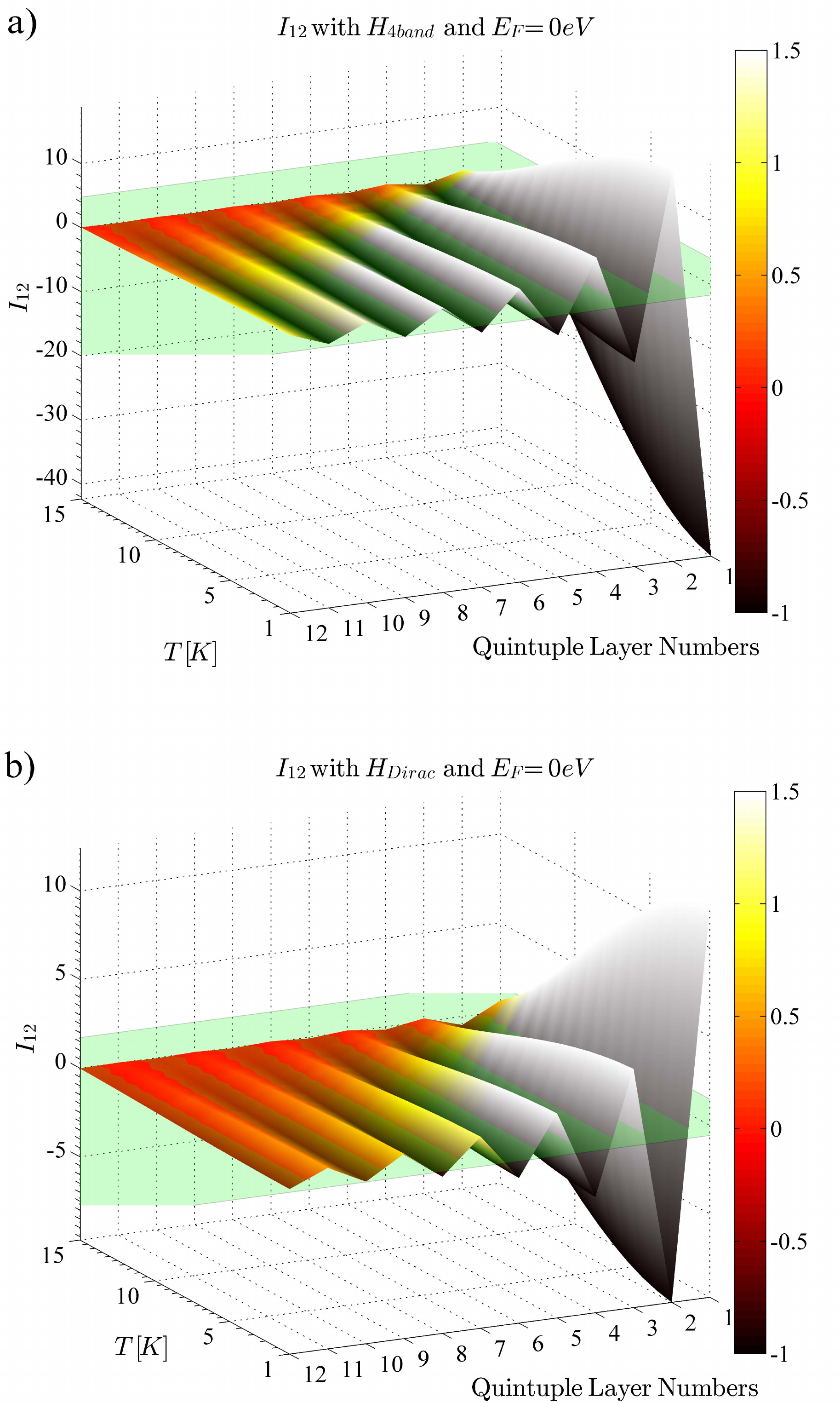}
\caption{The interlayer exchange coupling constant $I_{12}$ as a function of temperature and number of QL, with 4-band Hamiltonian (a) and Massive Dirac Hamiltonian (b). The oscillating ferromagnetic ($I_{12}<0$) – antiferromagnetic ($I_{12}>0$) coupling behavior are shown in both cases, but with a sign change.}
\label{fig2}
\end{figure}

%%%%%%%%%%%%%%%%%%%%%%%%%%%%%%%%%%%%%%%%%%%%%%%%%%%%%%%%%%%%%%%%%%%%%%%%%%%%%%%%%%%%%%%%%%%%%%%%%%%%%%%%%%%%%
\subsection{Effective Hamiltonian for Massive Dirac Fermion Surface States }
Contrary to the 4-band model which describes the bulk highest valence and lowest conduction states of Bi$_2$Se$_3$, the surface states are ideally only localized on the TI surface. However, due to the band bending effect which allows surface-state confinement near the interface, multiple surface states penetrate into the bulk \cite{44bahramy2012emergent}, including the Dirac surface states, M-shape valence states and Rashba-split conduction states. The strong band bending effect in Bi$_2$Se$_3$ can result in a deep penetration of states $\sim$12 QL. Thus, for thin TI spacer, it is still important to consider the possibility that the surface states participating in the interlayer magnetic coupling.

\begin{figure*}
\centering
\includegraphics[width=2.1\columnwidth]{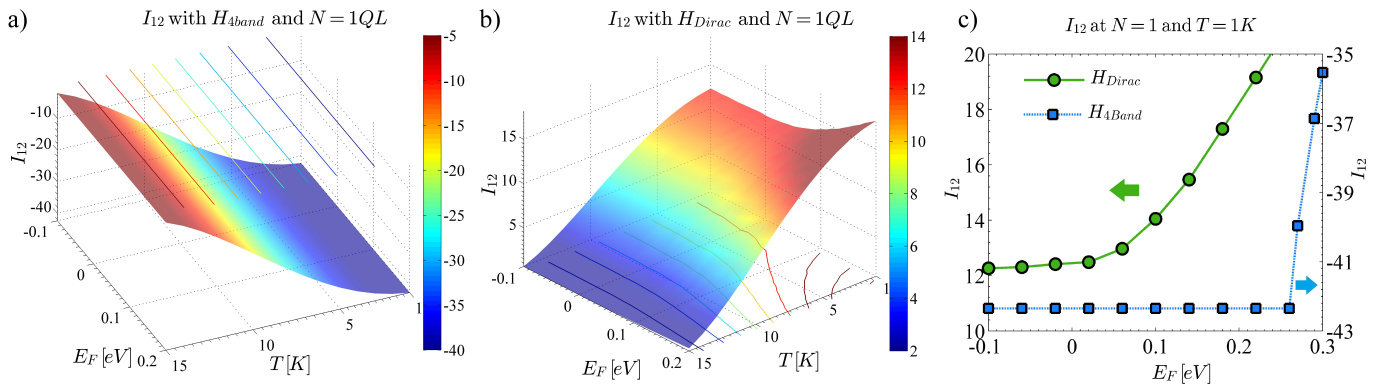}
\caption{a-b Interlayer coupling constant $I_{12}$ at $T=\unit[1]{K}$, as a function of temperature and Fermi level, for the 4-band and Dirac Hamiltonian, respectively. Within the bulk-band gap, $I_{12}$ does not change for the 4-band Hamiltonian, while $I_{12}$ is sensitive to $E_F$ for the Dirac Hamiltonian. c) The comparison of interlayer coupling constant $I_{12}$ between 4-band Hamiltonian and Dirac Hamiltonian, at 1 QL and $\unit[1]{K}$. We see that for the 4-band Hamiltonian $I_{12}$ remains constant while for Dirac Hamiltonian $I_{12}$ keeps changing. This fact can be used to identify the TI states participating in the proximity effect. We can also see that above the bulk band gap, the 4-band $I_{12}$ starts to change dramatically.}
\label{fig3}
\end{figure*}

For the purpose of qualitative demonstration, we neglect the M-shape valence states and Rashba-split conduction states, but only keep the Dirac states. The effective Hamiltonian for the Dirac states with gap opening can be written as\cite{1luo2013massive}

\begin{eqnarray}\label{eq9}
   {{H}_{2D}}(k)=D{{k}^{2}}I+
  \left( \begin{matrix}
   \mathcal{H}_{D}\!+\!M\cdot \sigma  & tI  \\
   tI &\!\! -\mathcal{H}_{D}\!+\!M\cdot \sigma   \\
\end{matrix} \right)
\end{eqnarray}

where $\mathcal{H}_{D}=\hbar {{v}_{F}}({{\sigma }_{x}}{{k}_{y}}\!-\!{{\sigma }_{y}}{{k}_{x}})$.
For 4 QL Bi$_2$Se$_3$ and magnet MnSe, $M=\unit[28.2]{meV}$, $t=\unit[17.6]{meV}$, $D=\unit[0.098]{eV\cdot nm^{2}}$, ${{v}_{F}}=\unit[2.66\times 10^5]{m/s}$. For simplicity we keep these parameters fixed when varying the thickness of TI and the type of magnet.
The eigenvalues can be written as

\begin{eqnarray}\label{eq10}
   &&\quad\quad E(k)=D{{k}^{2}}\pm\\
   &&\sqrt{\!{{\hbar }^{2}}v_{F}^{2}{{k}^{2}}\!\!+\!\!{{M}^{2}}\!\!+\!\!{{t}^{2}}\!\!+\!\!2\!\sqrt{\!\!{{M}^{2}}{{t}^{2}}\!+\!{{\hbar }^{2}}v_{F}^{2}{{({{M}_{x}}{{k}_{y}}\!\!-\!\!{{M}_{y}}{{k}_{x}})}^{2}}}}\nonumber
\end{eqnarray}

In sum, the interlayer coupling constant $I_{12}$ can be thus be calculated by substituting eq. (\ref{eq6}) back to eq. (\ref{eq2}). The eigenvalues in eq. (\ref{eq10}) and eq. (\ref{eq8}) can be used to obtain the magnetic susceptibility based on eq. (\ref{eq4}). From the modeled Hamiltonian approach, since we are more interested in a qualitative behavior rather than a quantitative magnitude, we regard eq. (\ref{eq5}) valid at finite $q$ values to incorporate the orbital contribution.

%%%%%%%%%%%%%%%%%%%%%%%%%%%%%%%%%%%%%%%%%%%%%%%%%%%%%%%%%%%%%%%%%%%%%%%%%%%%%%%%%%%%%%%%%%%%%%%%%%%%%%%%%%%%%
\section{Results and Discussions}
The interlayer coupling constant $I_{12}$ as a function of QL number and temperature are shown in Fig. \ref{fig2}, using the bulk 4-band Hamiltonian (eq. \ref{eq7}, Fig. \ref{fig2}a) and Dirac Hamiltonian (eq. \ref{eq9}, Fig. 2b). It is remarkable to see that at the same QL number, a sign difference of $I_{12}$ exists when the interlayer coupling are contributed by the valence and conduction electrons or Massive Dirac electrons. This is not only physically reasonable, due to the diamagnetic nature of bulk Bi$_2$Se$_3$ and paramagnetic nature of the surface states, but also agrees with the recent experimental report [45] which is able to extract paramagnetic Dirac susceptibility in the diamagnetic background in TI.

The significance of the sign difference can hardly be overestimated. In device application using TI/FMI proximity effect, it requires the exchange coupling of FMI with the Dirac surface states to open up the surface band gap. However, the FMI may also couple with other TI states simultaneously. Thus, the sign of $I_{12}$ would tell directly which TI states would dominate the proximity exchange coupling, and provide guidelines to suppress the proximity effect with other TI states while keeping the Dirac surface states dominant for future device design.  Moreover, with the aid of external magnetic field, it is theoretically possible to resolve the relative weights of coupling strength from TI Dirac state and other states, since they have different responses to external magnetic field.

Besides the sign change, the dependence of Fermi level provides further evidence to identify the TI states involved in the proximity effect (Figs. \ref{fig3}a and b). We see that for the bulk $p_z$ bands (Fig. \ref{fig3}a), $I_{12}$ is insensitive with Fermi level $E_F$ within $\sim\unit[0.3]{eV}$ bulk band gap (Fig. \ref{fig3}c, blue square curve), whereas on the contrary, a sensitive change of $I_{12}$  with $E_F$ (Fig. \ref{fig3}b and Fig. \ref{fig3}c, green circle curve) is shown when coupled with the Dirac states. Therefore, by varying the Fermi level and measuring the variation of $I_{12}$, it is in principle  possible to resolve the particular TI states contributing to the exchange coupling, and determine the relative weights in the proximity effect as well.

Due to the oscillating coupling behavior of nearby QL number, the thickness fluctuation becomes one factor which makes the resulting $I_{12}$ deviated from the ideal case, and hinders further extraction to determine the weights in the proximity effect. In the condition that the lateral correlation length is large enough $\xi >z$,  the averaged effect coupling constant $\bar{I}_{1,2}$ can be written as averaging over thickness fluctuations \cite{29levy1998range}

\begin{equation}\label{eq11}
{{\bar{I}}_{1,2}}=\int{dzP(z)I(z)}
\end{equation}
where $P(z)$ is the distribution function of spacer thickness. For simplicity we define
\begin{equation}\label{eq12}
P(z)=\frac{1}{\sqrt{2\pi }\sigma }\exp \left( -\frac{{{(z-\bar{z})}^{2}}}{2{{\sigma }^{2}}} \right)
\end{equation}

\begin{figure}
\begin{center}
\includegraphics[width=0.48\textwidth]{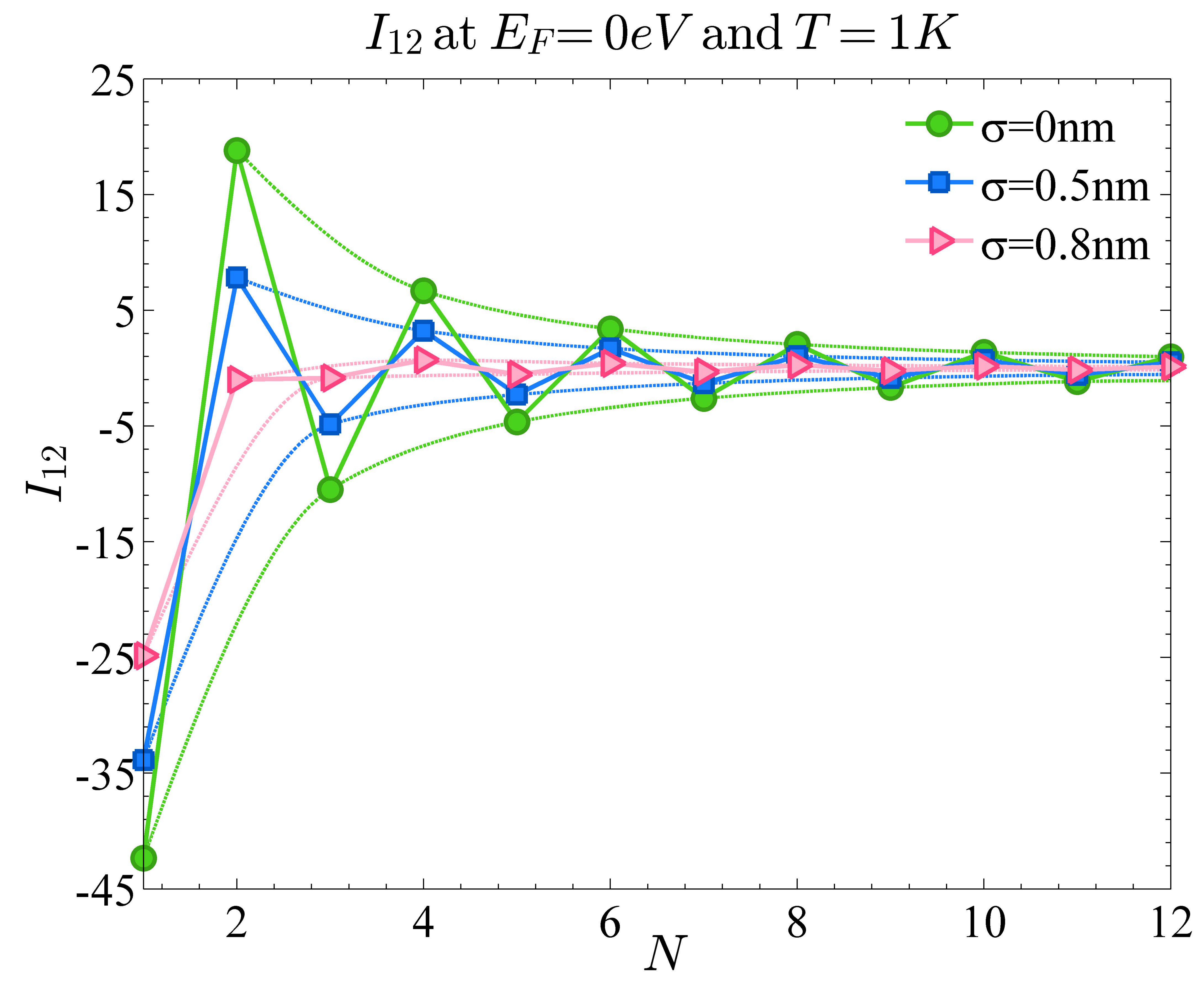}
\end{center}
\caption{The interlayer coupling constant $I_{12}$ at various thickness fluctuation , $\sigma=$ 0.5 and $\unit[0.8]{nm}$, using a 4-band Hamiltonian model at $E_F=\unit[0]{eV}$ and $T=\unit[1]{K}$. Stronger thickness fluctuation has a smoothing effect on the overall coupling constant, and may hamper the manifestation of TI states participating in the proximity effect. }
\label{fig4}
\end{figure}
where $\sigma$ is the thickness variation. Since the Se-Bi-Se-Bi-Se atomic layers within 1 QL is the strong chemical bonding, while the bonding between QLs is weaker van der Waals interaction, we still use 1 QL as the unit of thickness and discretizing $z\propto d$ with $d$ the thickness of 1 QL, without considering the possibility to break the chemical bonds within 1 QL which leads to fractional thickness in the unit of 1 QL. However, $\sigma$ can still be arbitrary as it denotes the relative weights for different thicknesses to appear in the layered structure. As an illustration, the resulting change of $I_{12}$ for 4-band Hamiltonian with different $\sigma$ are shown in Fig. \ref{fig4}. When the thickness fluctuation increases, the resulting averaged $\bar{I}_{1,2}$ drops dramatically. Thus, in order to determine the particular TI states involved in the proximity exchange coupling as well as their relative weights, high-quality samples with negligible thickness fluctuation are desirable.

%%%%%%%%%%%%%%%%%%%%%%%%%%%%%%%%%%%%%%%%%%%%%%%%%%%%%%%%%%%%%%%%%%%%%%%%%%%%%%%%%%%%%%%%%%%%%%%%%%%%%%%%%%%%%
\section{CONCLUSIONS}
We have provided a systematic approach to illustrate the feasibility that how interlayer exchange coupling in FMI/TI/FMI structure can help understand the TI/FMI proximity effect, with the capability to identify the TI states involved in the proximity exchange. By changing the external magnetic field or Fermi level, the weights for the exchange coupling between FMI and the desired TI Dirac states can be obtained. Such information can hardly be obtained directly by the experimental probes such as ARPES, MOKE, PNR or transport, since this approach circumvents the complications of the TI-FMI interaction, but infers the TI states from the simpler indirect FMI-FMI coupling using TI states as medium. In this perspective, the interlayer coupling between two FMIs in the FMI/spacer/FMI structure is not only an interesting phenomenon by itself, but also can be regarded as a probe to study the properties of the spacer.

Moreover, since the interlayer exchange coupling in magnetic multilayers, such as Fe/Cr superlattice\cite{49BaibichGMR}, has played a significant role in giant magnetoresistance (GMR) \cite{46Fert19951,47Parkin1990GMR,48Parkin1991GMR}, the present work also sheds light on the application of magnetic data storage and magnetic field sensors. As shown in Fig. \ref{fig3}c, the interlayer exchange coupling constant could be tuned when coupling with Dirac states of TI. This provides a method to achieve electrically controlled magnetic coupling, with reversibility due to gating, and rapid response due to high-mobility, backscattering-protected Dirac electrons. The only prerequisite is that the exchange coupling with massive Dirac states should overcome the bulk TI states. This may be realized in thinner TI film where the bulk bands diminish whereas the surface bands dominate. Therefore, further studies of interlayer exchange coupling in TI based magnetic layers for GMR applications, are highly desired.

\acknowledgements
M.L. would thank Prof. Ju Li for generous support and helpful discussions. J.S.M. and F.K. would like to thank support from the MIT MRSEC through the MRSEC Program of the NSF under award number DMR-0819762. J.S.M. would like to thank support from NSF DMR grants 1207469 and ONR grant N00014-13-1-0301.
%\appendix
%\section{}\label{appendix1}
\bibliography{Bibliography}
\end{document}